\RequirePackage{fix-cm}

 \documentclass[showkeys,preprint]{revtex4}          
\usepackage{graphicx}
\usepackage{amssymb}
\usepackage{amsbsy}
\usepackage{ulem}
\usepackage{textcomp}
\usepackage{hyperref}
\usepackage[hang,small]{caption}
\usepackage{amssymb}
\usepackage{amsmath}
\usepackage{diagbox}
\usepackage{fancyhdr}
\usepackage{subfigure}
\usepackage{color}
\usepackage{enumitem}
\usepackage{cases}

\usepackage{comment} 

\usepackage[toc,page]{appendix} 
\usepackage{array} 

\begin{document}

\title[]{Influence of morphological parameters in 3D composite materials on their effective thermal properties and comparison with effective mechanical properties
}

\author{Sophie Lemaitre{\textsuperscript{a}}}
\author{Vladimir Salnikov{\textsuperscript{b}}} 
\author{Daniel Cho\"i{\textsuperscript{c}}}
\author{Philippe Karamian{\textsuperscript{d}}}

\affiliation{a,c,d. Nicolas Oresme Mathematics Laboratory \\ University of Caen Normandy, \\ 
  CS 14032, Bd. Mar\'echal Juin,  BP 5186\\
  14032, Caen Cedex,  France \\b. University of Luxembourg\\ 
  \email{a. sophie.lemaitre@unicaen.fr,b. vladimir.salnikov@uni.lu,c. daniel.choi@unicaen.fr, d. philippe.karamian@unicaen.fr}
}

\begin{abstract}
 In this paper we study the effective thermal behaviour of 3D representative volume elements (RVEs) of two-phased composite materials constituted by a matrix with cylindrical and spherical inclusions distributed randomly, with periodic boundaries.
Variations around the shape of inclusions have been taken into account, by corrugating shapes, excavating and/or by removing pieces of inclusions. The effective behaviour is computed with the help of  homogenization process based on an accelerated FFT-scheme 
giving the thermal conductivity tensor. Several morphological parameters are also taken into account for instance the number and the volume fraction of each type of inclusions,... in order to analyse the behaviour of the composite for a large number of geometries. We compare the results obtained for RVEs  with and without variations, and then with the mechanical results of such composite studied in our previous paper.

\keywords{
Composite material / Cylindrical and spherical reinforcements / Mechanical and thermal properties / Stochastic homogenization / FFT}
\end{abstract}
\maketitle
\section{Introduction and motivations}
\indent 

The study of composite materials is of great interest in various  fields  such as aerospace engineering and industry in which the metallic materials can be advantageously replaced. To that purpose, the modelling and numerical simulations of such composites are essential to the design with optimal electrical, mechanical and thermal properties, since  authentic experiments are hugely expensive.

\begin{figure}[H]
  \centering
  \subfigure[First variation: corrugation]
{\label{def1}\includegraphics[scale=0.2]{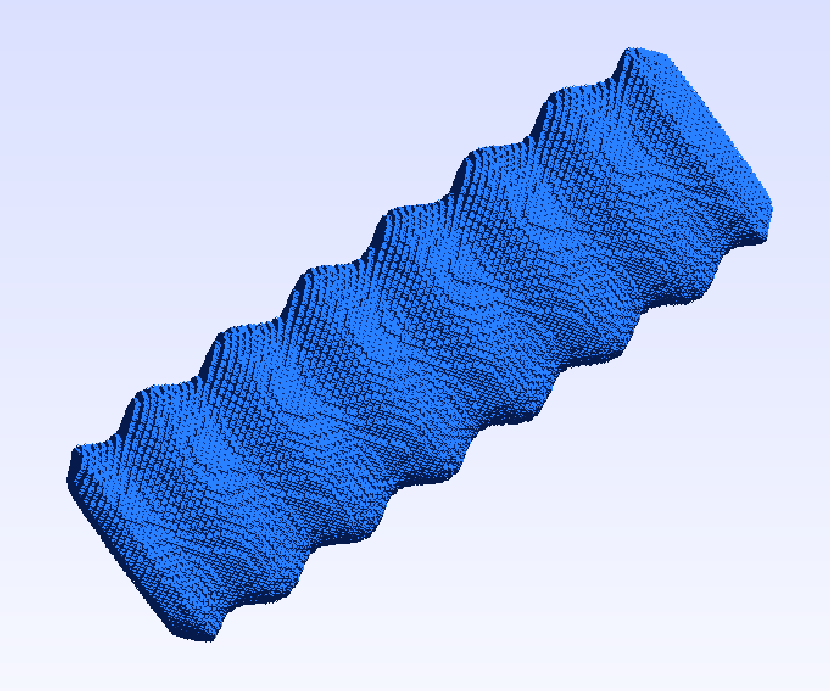}}  
  \hspace{5pt}
  \subfigure[Second variation: broken piece]
  {\label{def2}\includegraphics[scale=0.2]{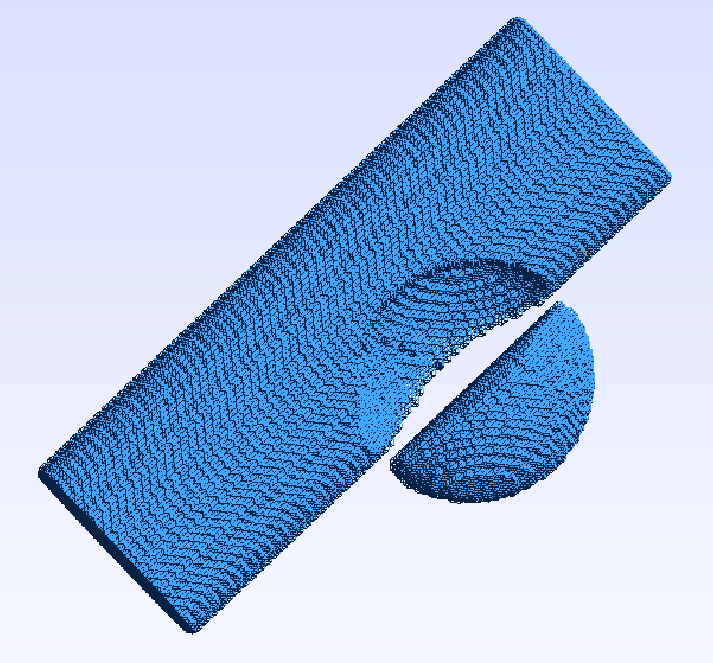}}
  \caption{The two types of variations studied}
  \label{defect}
\end{figure}

In this paper, composite materials constituted by a  matrix (modelling classical polymers or plastics) reinforced by poorly or highly conductive inclusions are considerated. The inclusions are constituted by a mixture of cylinders and spheres. We also consider two kinds of variations around the shape of inclusions : the spheres and cylinders which are corrugated or some pieces of the inclusions are excavated and/or removed to produce irregular shapes as it is shown in Fig. \ref{defect}.

This study, focuses on  the influence of the morphology and the influence of the variation of shapes for the homogenized thermal conductivity of such composite material by extending  the study of the mechanical behaviour published in a previous paper (\cite{Salnikov1}) to thermal properties. The main interest of this study is to give two information about a material, for instance its mechanical and thermal behaviour.

\begin{figure}[H]
  \centering
  \includegraphics[scale=0.6]{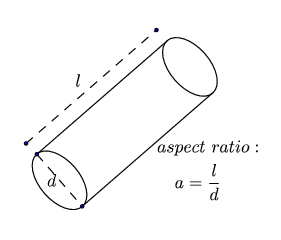}
  \caption{Definition of the aspect ratio for a cylinder}
  \label{aspect}
\end{figure}

For the influence of the morphology, our work has consisted in the study of the impact of the geometry and the volume fraction of globular and fibre type inclusions. For this purpose, we have taken into account several parameters such as the number, the volume fraction of each type of inclusions, and for the fibre type inclusions, we have studied the aspect ratio (Fig. \ref{aspect}) which are able to distinguish short or long fibre.

For the influence of the variations of shape inclusions (Fig. \ref{defect}), we have varied  the wave parameter for the first variation (see Fig. \ref{wave}) and the volume fraction of defects ( $f_{def}$) which designates the volume fraction of pieces of inclusions excavated for the second variation.

Classically (\cite{segurado}, \cite{ghossein}, \cite{zhao}), \cite{man}, the key idea is to consider a large enough sample of a composite material and compute the macroscopic parameters such as the components of the conductivity tensor in the case of isotropic or quasi-isotropic materials. Such samples are called Representative Volume Element (RVE).

For the same reasons described in our previous work, we use a series of samples (RVEs) generated randomly (\cite{Salnikov2}), by controlling their parameters such as the number of spheres and cylinders, the aspect ratio, the volume fraction and we perform the computation for each of them before averaging the result.

The macroscopic thermal properties computations of the RVEs are obtained by  homogenization methods based on  FFT techniques which had been  introduced in \cite{Eyre and Milton}, \cite{Moulinec1} and \cite{ghossein} that  we have adapted for thermal analysis. The RVEs are modelled as large 3D images (over 7 million voxels per image). Finite element method also can be used, but we choose the FFT-based method due to the  computational efficiency considerations. We used finite element method mainly for comparison and validation of our FFT-based code.

To compare the mechanical and thermal behaviour, we follow the presentation introduced in (\cite{Salnikov1}) and we exhibit the trends with the same parameters which have been shown in our previous paper (\cite{Salnikov1}) and which are more significant.

The paper is organized as follows. The section 2 is devoted to the presentation of the sample generation and the computational techniques. In section 3, we present the results of the computed thermal properties of considered composite materials. Finally, in section 4, we compare the trends observed with thermal properties  with the mechanical ones and we give our conclusions and outlooks.

In the appendices, we recall the thermal model used and the FFT-based homogenization method. In the last appendix, to help the reader a table with all parameters and their range is also given .

\section{RVE generation and computational techniques}
\indent 


\subsection{RVE generation}
\indent 

In composite homogenization, the RVE generation including inclusions is a key stage and a real challenge especially with periodic boundary conditions. The generation scheme must be robust, reliable, fast and able to reach high volume fraction of inclusions.
The RVE consists of a random mixture in 3D of inclusions which are spherical and cylindrical without intersection in a first state, and in a second state with variations on inclusions shapes. There are many works with ellipsoidal or spherical inclusions (\cite{segurado}, \cite{ghossein},\cite{zhao},\cite{man}) but the case with cylinders presents geometric difficulties that we have resolved and detailed in \cite{Salnikov2}. However, another key point is how to generate samples randomly. 
There are two  families of  generation algorithm: Random Sequential Adsorption (RSA) and Molecular Dynamics (MD).
RSA is based on sequential addition of inclusion until there is no intersection with any other inclusion already inserted. This method generates RVEs with about $30\%$ volume fraction and aspect ratio (see \ref{aspect}) smaller than 16. For higher volume fraction and for high aspect ratio, this method can stuck and one prefers the second method based on molecular dynamics. MD consists in generating all inclusions and after moving them to the desired configuration using a time-driven version.
This method allows the generation up to $50\% - 60\%$. For more details, the two algorithms are given in \cite{Salnikov2}. Our calculations have been made with the first method due to the choice of the morphology parameters. To illustrate the type of RVE used, we present two examples with a mixture of non intersecting spherical and cylindrical inclusions with low and high volume fraction.

\begin{figure}[H]
  \centering
  \subfigure[$f_{cyl}=f_{sp}=0.03$]{\label{fig1a}\includegraphics[scale=0.2]{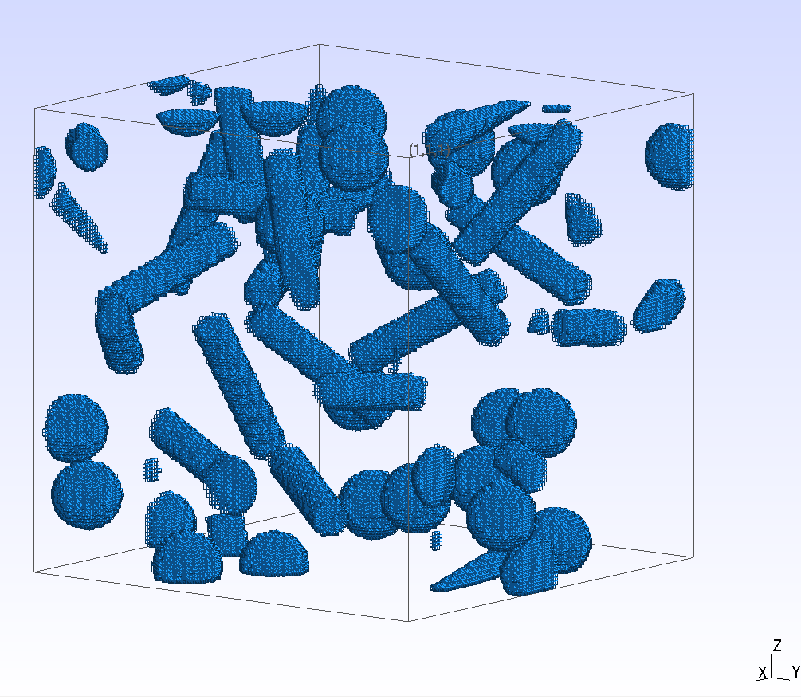}}
  \hspace{5pt}
  \subfigure[$f_{cyl}=f_{sp}=0.15$]{\label{fig1b}\includegraphics[scale=0.2]{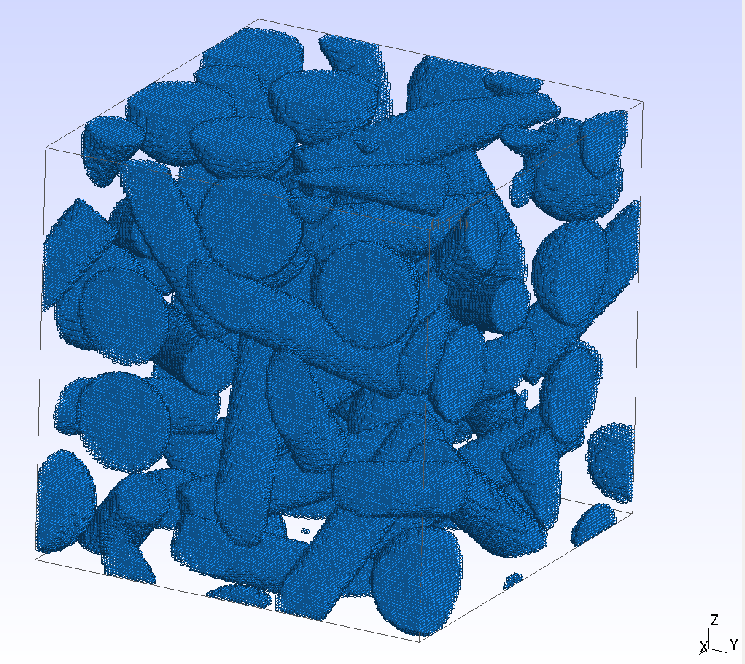}}
  \caption{3D views of generated samples for low and high volume fraction with cylinders and spheres without intersection and boundary conditions ($n_{cyl}=n_{sp}=20$)}
  \label{fig1}
\end{figure}

Concerning the inclusions with shape variations (\cite{CSMA}), we consider corrugated  inclusions, or, we remove some pieces of  inclusions (see the Fig. \ref{defect}) after generating spheres and cylinders at the pixelization (or voxelization) step. For these RVEs, we have created inclusions with distorted shapes to compare their behaviour with the non distorted ones.
For our computations, we use a "wave parameter" between 0 and 1 (see Fig. \ref{wave}) in order to corrugate the inclusions and a volume fraction of defects ($f_{def}$)between 0 and 0.27.

\begin{minipage}{0.4\linewidth}
  \includegraphics[width=\linewidth ,trim=30 20 100 20,clip=true]{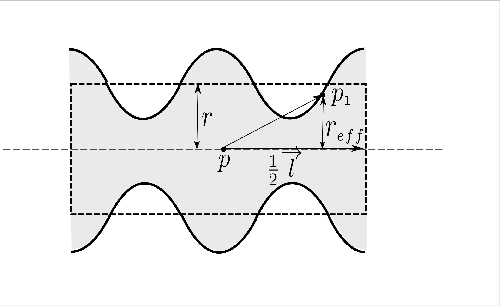}
  \label{wave}
  Fig.\ref{wave}: 2D cylinder view with 3 corrugations
\end{minipage}\hfill
\begin{minipage}{0.6\linewidth}
\hspace{0.5cm} Legend of Fig. \ref{wave}: 
\begin{enumerate}
\item[-] $r$ is the cylinder radius,
\item[-] $r_{eff}$ is the corrugated radius.
$$ r_{eff}= r \left(1+ wave \times f(p_1)\right)$$
\item[] where $f$ is a periodic function.
\end{enumerate}
\end{minipage}

\begin{figure}[H]
  \centering
  \subfigure[$f_{cyl}=f_{sp}=0.06,wave=0.12$]{\label{fig2a}\includegraphics[height=5cm]
  {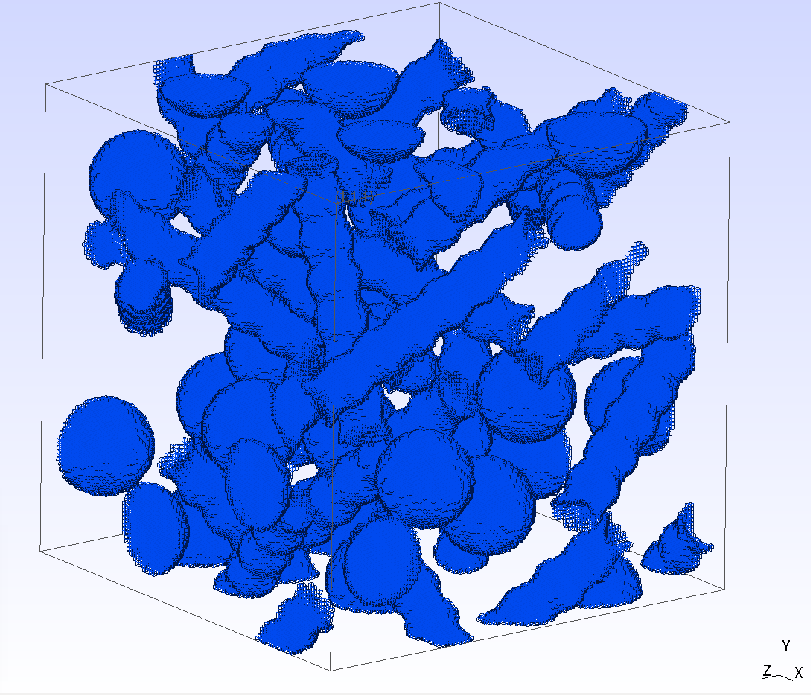}}
  \hspace{5pt}
  \subfigure[$f_{cyl}=f_{sp}=0.09$, $f_{def}=0.1 $]{\label{fig2b}\includegraphics[height=5cm]{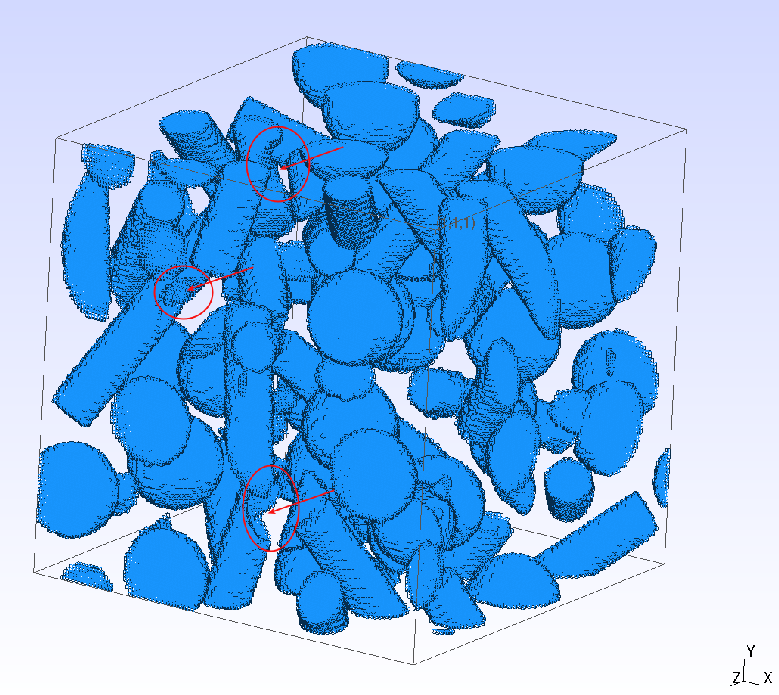}}
  \caption{3D views of generated samples with cylinders, spheres, and irregular shapes ($n_{cyl}=n_{sp}=20$) (Examples of pieces taken apart in red) }
  \label{fig2}
\end{figure}

\subsection{FFT-based homogenization scheme to study thermal properties}
\indent 

This section describes briefly how the thermal behaviour of a composite material is obtained, we let the reader see the appendices for more details. Once the RVEs are generated, we can proceed to the computation of effective thermal properties. Then, to evaluate these properties, a stochastic homogenization is done (see appendix \ref{approach}).

The thermal model used for our computation is  more precisely described in appendix \ref{model}. 

This approach has a twofold advantage, both in time and memory consumption in comparison to the finite elements method. Furthermore, there are several FFT-based numerical schemes such as the 'basic scheme', the 'dual scheme', the augmented Lagrangian scheme, the 'polarization scheme' and the 'accelerated scheme'. The last algorithm has caught our attention for its computational efficiency same as in the elastic case (see \cite{Salnikov2}) when the contrasts of the materials studied are low or high but never infinite, exactly as in the framework described by Moulinec and Silva in \cite{Moulinec and Silva}. In this paper, the authors have compared the different schemes and have proved that the accelerated scheme is the optimal compromise (except for composite material with infinite contrast).
 
All these algorithms need to be initialized with a reference tensor which depends on the algorithm chosen. The materials studied here are isotropic with a conductivity tensor of the form $kI$ ($k$ positive real). For the accelerated scheme, according to \cite{Eyre and Milton} and \cite{Moulinec and Suquet 2001} the constant reference thermal tensor $\Lambda^{0}$ is set to:
 $$\Lambda^{0}=-\sqrt{\underset{x \in \{1,c\}}{min(x)} \times\underset{x \in \{1,c\}}{max(x)} } \: I=-\sqrt{c} \:I,$$
 where c is the contrast between two phases.
 
In our stochastic approach, several generations of samples are considered using same morphological parameters as the numbers and volume fractions of each type of inclusion (spheres and cylinders), the morphology of the cylinders with the aspect ratio (the ratio between the length and the diameter of the cylinder (Fig. \ref{aspect})). With each generated RVE, we are able to compute each homogenized conductivity tensor. We compute the thermal apparent conductivity coefficient as:
$$\lambda_{app}=\frac{1}{n}\sum_{i=1}^{n} \frac{1}{3}Tr(\Lambda^{i,hom})=\frac{1}{n}\sum_{i=1}^{n}\frac{1}{3} (\Lambda_{11}^{i,hom}+\Lambda_{22}^{i,hom}+\Lambda_{33}^{i,hom})$$
where:

$n$ is the number of RVEs with the same set of morphological parameters that is the number of runs,

$\Lambda^{i,hom}$ is the homogenized thermal conductivity tensor for the RVE i.\\

This approach is not new, and to determine how many runs are necessary, the standard deviation of the set of samples is calculated and one can see that 10 (sometimes 20) runs are enough. Taking the average, the apparent thermal conductivity denoted by $\lambda_{app}$ is obtained.

\section{Thermal properties of composite materials}
\indent

This section is devoted to the presentation of the results of our numerous computations. We have varied the morphological parameters (number of inclusions, volume fraction, aspect ratio, defects, corrugations) with an adapted range according to the influence observed. The first results are some tests which allow us to validate the methods, giving expected results. Then, combinations of morphological parameters are tested and give the following results.

In order to compare with the mechanical behaviour, we use the same RVEs generated for the mechanical study and for the same reasons mentioned in \cite{Salnikov1}, we work with spheres and cylinders that represent globular inclusions and microfibre reinforcements. We recall here the notations  for each generated sample, $f_{sp}$ and $f_{cyl}$ are respectively the spheres and cylinders volume fraction, $n_{sp}$ and $n_{cyl}$ are respectively the spheres and cylinders number inside the RVE and for a cylinder we need another parameter: the aspect ratio which is the ratio between the length and the diameter of the cylinder (Fig. \ref{aspect}). In the appendix \ref{notations}, a table resumes the notations used and the range for each series of computations.

\subsection{Validation tests}\label{basic}
\indent 

As explained in \cite{Salnikov1}, we have analysed the dependence of effective properties on the number of inclusions with all the other parameters fixed. We have supposed materials (matrix and inclusions) isotropic and homogeneous. We have observed that off-diagonal terms of the homogenized thermal conductivity tensor approach zero or are not preponderant. Then, the average of the three diagonal terms for each RVE is taken for the effective thermal properties and the average for the set of geometries generated with the same parameters is taken to obtain the value of $\lambda_{app}$. In general 10 RVEs have been generated for each set of parameters. We have represented this value of  $\lambda_{app}$ in the $y$ axis on all graphics. 

\begin{figure}[H]
  \centering
  \includegraphics[scale=0.45]{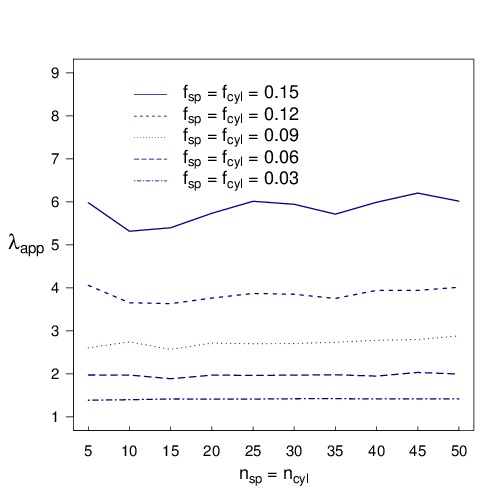}
  \caption{The dependence of thermal conductivity of a composite material on the number of inclusions for the volume fraction being fixed with a contrast of 2048 and an aspect ratio = 5.}
  \label{fig3}
\end{figure}

In Fig. \ref{fig3},  the volume faction of cylinders and spheres are equal and allow us to obtain RVEs with a total volume fraction from $0.06$ to $0.30$. The more interesting results are observed for a large contrast. Fig. \ref{fig3} shows that from the number of 20 that is from 20 cylinders and 20 spheres, $\lambda_{app}$ stabilizes though for higher volume fraction, the effect is more pronounced like for mechanical properties. 

\begin{figure}[H]
  \centering
  \includegraphics[scale=0.45]{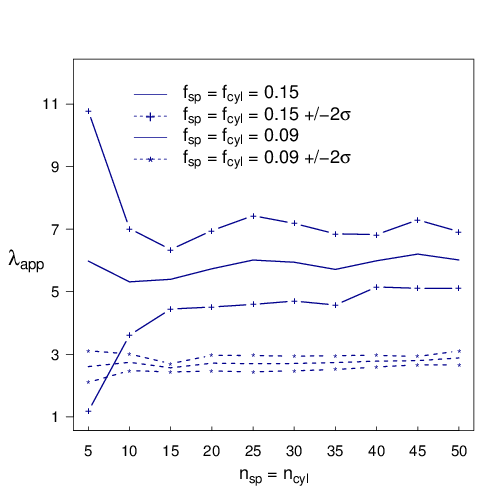}
  \caption{$f_{sp}=f_{cyl}=0.15$ and $f_{sp}=f_{cyl}=0.09$, contrast = 2048 and aspect ratio = 5.}
  \label{fig3abis}
\end{figure}

\begin{figure}[H]
  \centering
  \includegraphics[scale=0.45]{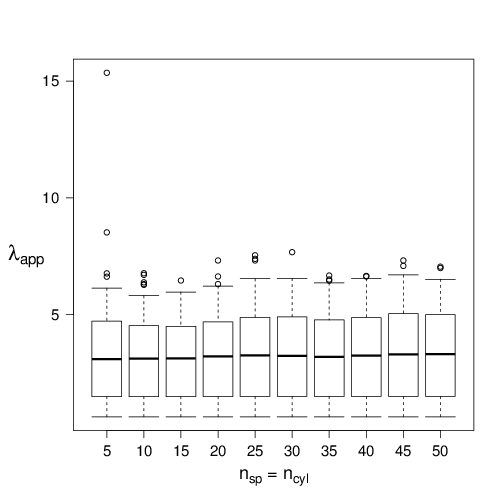}
\caption{The repartition with boxplots for $f_{sp}=f_{cyl}= 0.15$ and aspect ratio = 5.}
  \label{fig3bbis}
\end{figure}

In the FIG. \ref{fig3abis}, we have drawn the curves for 
$f_{sp}=f_{cyl}=0.15$ and $f_{sp}=f_{cyl}=0.09$ that is for a total volume fraction equal to $0.3$ and $0.18$ for each case, with curves which show the confidence interval with $\pm 2\sigma$. FIG. \ref{fig3abis} shows that the higher  the volume fraction of inclusions is, the higher  the dispersion is, and the higher  the number of inclusions is, the lower  the dispersion is. In the Fig. \ref{fig3bbis}, we have drawn the boxplots for each number of inclusion for the case $f_{sp}=f_{cyl}=0.15$. We note that the higher  the number of inclusions is and more stable the repartition of $\lambda_{app}$   is. We can note that from 20 (20 cylinders and 20 spheres), the range of $\lambda_{app}$ is also stable.
Thus for the further computations, a number of 20 inclusions is considered.

In the sequel, the study focuses on the influence of morphological parameters. We describe the trends for $\lambda_{app}$ versus the parameter studied.

\begin{figure}[H]
  \centering
  \subfigure[$n_{sp}=15$]{\label{fig4a}\includegraphics[scale=0.38]{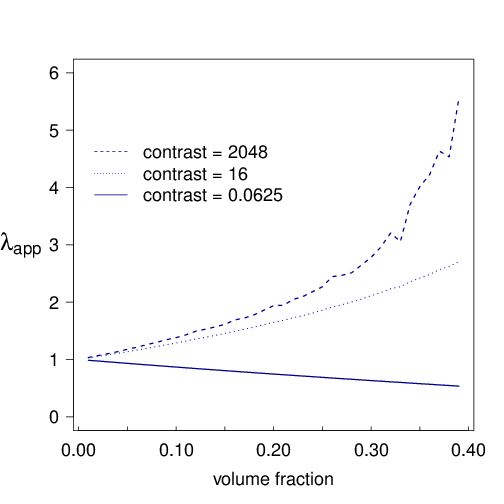}}
  \hspace{2pt}
  \subfigure[$n_{sp}=20 $]{\label{fig4b}\includegraphics[scale=0.38]{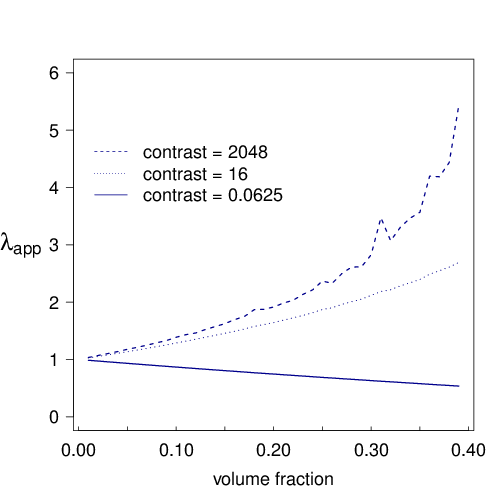}}
  \hspace{2pt}
  \subfigure[$n_{sp}=25 $]{\label{fig4c}\includegraphics[scale=0.38]{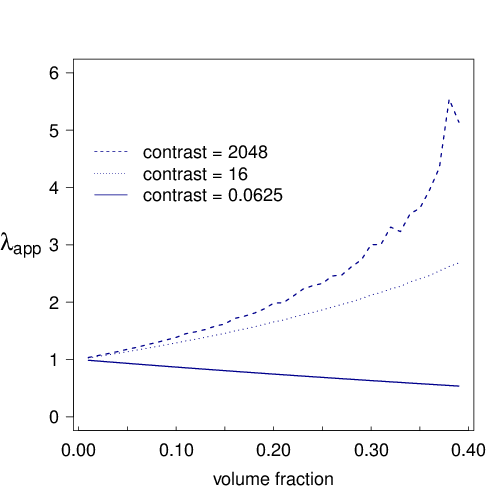}}
  \caption{Dependence of thermal conductivity of a composite material on the volume fraction of spherical inclusions,  $n_{sp}=15,20,25$.}
  \label{fig4}
\end{figure}

\begin{figure}[H]
  \centering
  \subfigure[Contrast = 0.0625]{\label{fig5a}\includegraphics[scale=0.7]{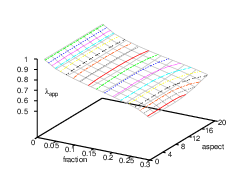}}
  \subfigure[Contrast = 16]{\label{fig5b}\includegraphics[scale=0.7]{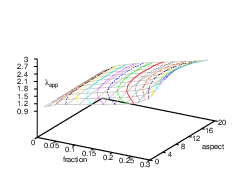}}
  \hspace{2pt}
  \subfigure[Contrast = 256]{\label{fig5c}\includegraphics[scale=0.7]{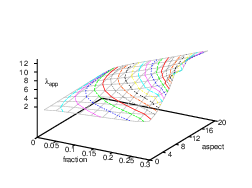}}
  \hspace{2pt}
  \subfigure[Contrast = 2048]{\label{fig5d}\includegraphics[scale=0.7]{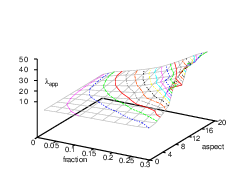}}
  \caption{The dependence of thermal conductivity of a composite material on the volume fraction and aspect ratio of cylindrical inclusions,  $n_{cyl}=20$.}
  \label{fig5}
\end{figure}

First, we only consider one type of inclusions: spherical inclusions (Fig. \ref{fig4}) and next cylindrical inclusions (Fig. \ref{fig5}).

Fig. \ref{fig4} shows that the trends are similar for 15, 20 or 25 spheres. We note that $\lambda_{app}$ increases non linearly for contrasts greater than 1 and decreases for contrasts smaller than 1. We also notice that these effects are more pronounced for contrast greater than 1. In Fig. \ref{fig5}, we observe the same conclusions more enhanced with adding the effect of the aspect ratio. Similar phenomenon had been noticed for the mechanical properties in \cite{Salnikov1}.

\subsection{Advanced morphology analysis}\label{advanced}
\indent 

This subsection is devoted to  the influence of morphology on the thermal conductivity. Namely,  the influence of morphological parameters (number, volume fraction, aspect ratio) on the apparent thermal conductivity is studied. 

We have already observed on Fig. \ref{fig5} that composite materials with cylindrical inclusions have larger apparent thermal conductivity when the aspect ratio is high. 
For the following tests, the number of cylinders has been fixed at 20 and we have studied the influence of both volume fraction of cylinders and their aspect ratio.

\begin{figure}[H]
  \centering
  \subfigure[Contrast = 0.0625]{\label{fig6a}\includegraphics[scale=0.35]{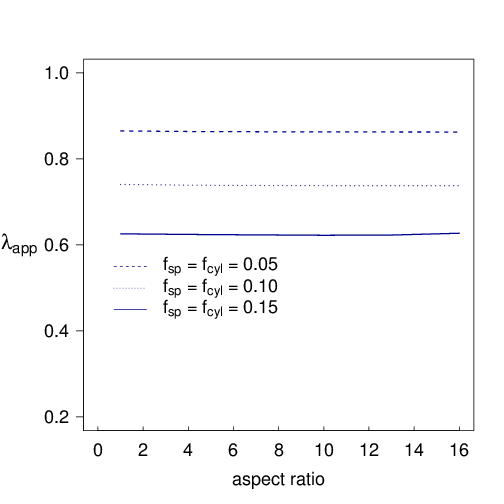}}
   \hspace{5pt}
  \subfigure[Contrast = 16]{\label{fig6b}\includegraphics[scale=0.35]{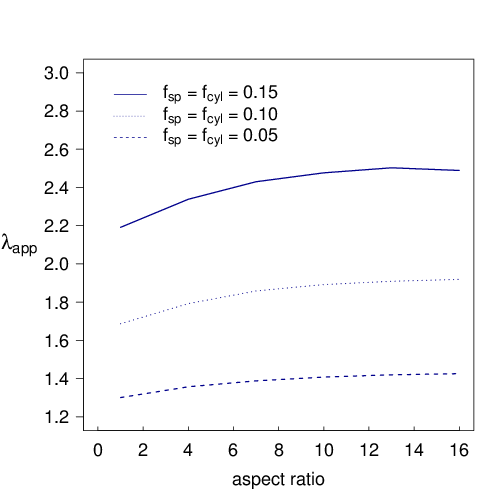}}
  \subfigure[Contrast = 256]{\label{fig6c}\includegraphics[scale=0.35]{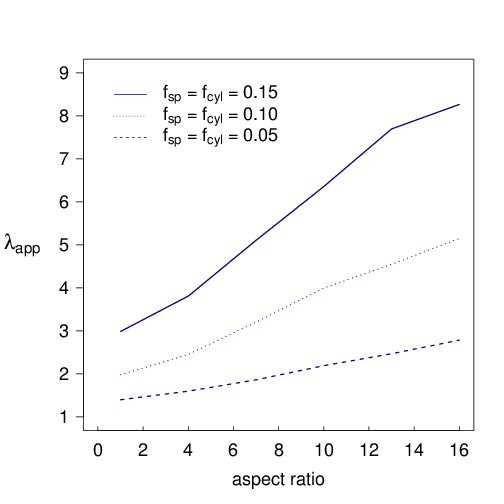}}
  \hspace{5pt}
  \subfigure[Contrast = 2048]{\label{fig6d}\includegraphics[scale=0.35]{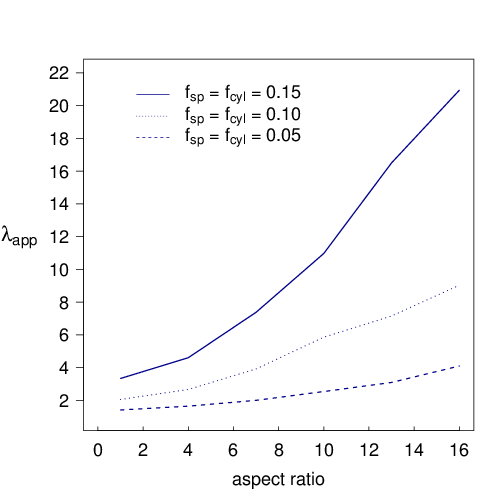}}
  \caption{The dependence of thermal conductivity of a composite material on the aspect ratio of the cylinders in the mixture of inclusions,  comparison for $n_{sp}= n_{cyl}=20$.}
  \label{fig6}
\end{figure}

In Fig. \ref{fig6}, we observe the behaviour of composite materials with a mixture of cylindrical and spherical inclusions. 
The number of each type of inclusions that is 20 spheres and 20 cylinders and the volume fraction being fixed for each computation, we have studied the influence of the aspect ratio.

We observe as in the mechanical case that for contrast greater than 1, the apparent thermal conductivity of the composite material increases even more with longer cylinders. Moreover, in a composite material with only cylindrical inclusions and contrast greater than 1, especially for higher contrast, the apparent thermal conductivity reaches higher values than in the case of a mixture of spherical and cylindrical inclusions. For smaller contrast (id smaller than 1), we remark that the presence or not of spherical inclusions does not change the apparent thermal conductivity compared to a material with only cylinders whereas we have observed that homogenized mechanical parameters increase in this situation.

\begin{figure}[H]
  \centering
  \subfigure[$f_{sp}= f_{cyl}=0.05$, contrast 0.0625]{\label{fig7a}\includegraphics[height=5cm]{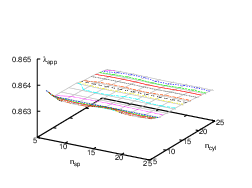}}
   \hspace{5pt}
  \subfigure[$f_{sp}= f_{cyl}=0.05$, contrast 2048]{\label{fig7b}\includegraphics[height=5cm]{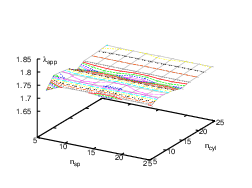}}
  \subfigure[$f_{sp}= f_{cyl}=0.1$, contrast 0.0625]{\label{fig7c}\includegraphics[height=5cm]{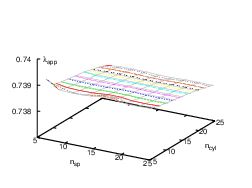}}
  \hspace{5pt}
  \subfigure[$f_{sp}= f_{cyl}=0.1$, contrast 2048]{\label{fig7d}\includegraphics[height=5cm]{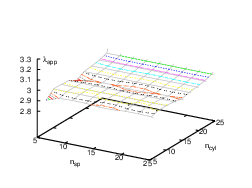}}
   \subfigure[$f_{sp}= f_{cyl}=0.15$, contrast 0.0625]{\label{fig7e}\includegraphics[height=5cm]{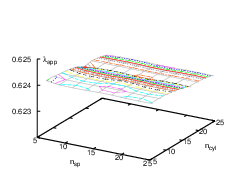}}
  \hspace{5pt}
  \subfigure[$f_{sp}= f_{cyl}=0.15$, contrast 2048]{\label{fig7f}\includegraphics[height=5cm]{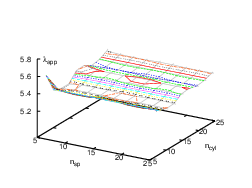}}
  \caption{The dependence of thermal conductivity of a composite material on the number of various inclusions for fixed volume fraction. Note the large difference between the scales.}
  \label{fig7}
\end{figure}

For the third series of tests on Fig. \ref{fig7}, we have fixed the volume fraction of each type of inclusions (spheres and cylinders). The aspect ratio of cylinders fixed at 5, we have studied the influence of both the number of spheres and the number of cylinders. 

We can observe that the influence of the number of spheres or cylinders whether for a contrast greater than 1 or less than 1, is negligible.

\begin{figure}[H]
  \centering
  \subfigure[Contrast 0.0625]{\label{fig8a}\includegraphics[height=5cm]{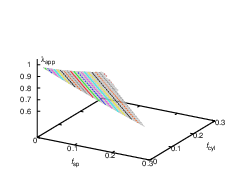}}
   \hspace{5pt}
  \subfigure[Contrast 0.0625]{\label{fig8b}\includegraphics[height=5cm]{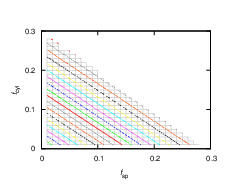}}
  \subfigure[Contrast 16]{\label{fig8c}\includegraphics[height=5cm]{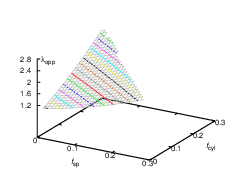}}
  \hspace{5pt}
  \subfigure[Contrast 16]{\label{fig8d}\includegraphics[height=5cm]{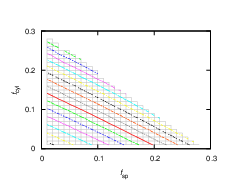}}
   \subfigure[Contrast 2048]{\label{fig8e}\includegraphics[height=5cm]{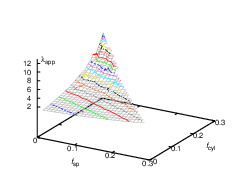}}
  \hspace{5pt}
  \subfigure[Contrast 2048]{\label{fig8f}\includegraphics[height=5cm]{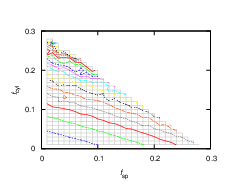}}
  \caption{The dependence of thermal conductivity of a composite material on the repartition of inclusions volume between spheres and cylinders. $n_{sp}= n_{cyl}=20$. 3D plot on the left and level sets map on the right.}
  \label{fig8}
\end{figure}

In the last series of tests on Fig. \ref{fig8}, we have studied the influence of volume fraction of spheres and cylinders. The number of cylinders is fixed at 20 and is equal to the number of spheres. The aspect ratio of cylinders is also fixed at 5. The plots of Fig. \ref{fig8} represent the effective thermal conductivity for a fixed sum of the volume fraction and the two types of inclusions. We observe that the apparent thermal conductivity increases rather than for high volume fraction of cylinders and high contrast.

\subsection{Variations around the shape of inclusions}
\indent 

The last two series of tests represent the case of inclusions with irregular shape. Let us recall that we have developed two algorithms (\cite{CSMA}) able to modify the shape of inclusions which are cylindrical or spherical. An example of each perturbation is shown in figure \ref{fig2a} . 

\begin{figure}[H]
  \centering
  \subfigure[Aspect ratio = 3]{\label{fig9a}\includegraphics[scale=0.35]{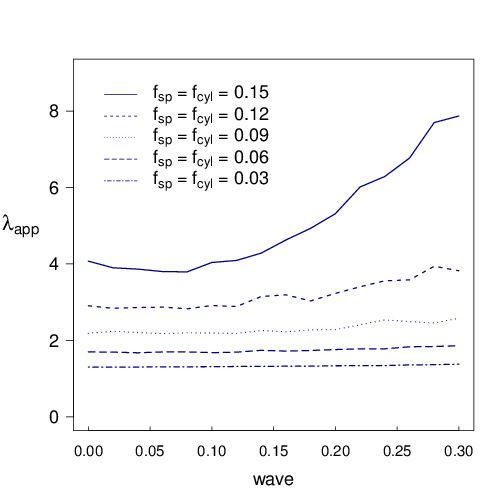}}
   \hspace{5pt}
  \subfigure[Aspect ratio = 6]{\label{fig9b}\includegraphics[scale=0.35]{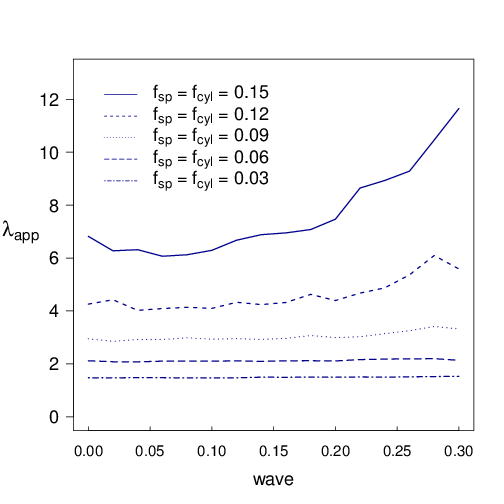}}
  \subfigure[Aspect ratio =  9]{\label{fig9c}\includegraphics[scale=0.35]{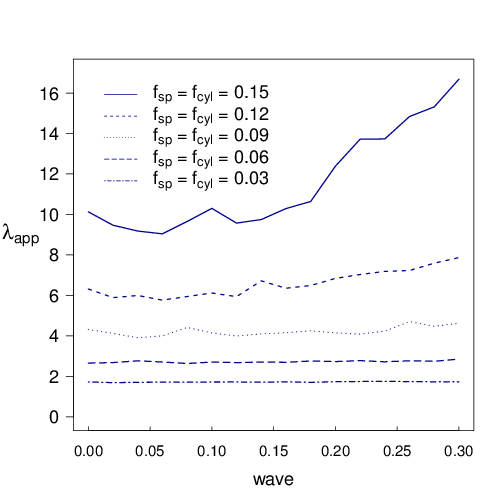}}
  \subfigure[Aspect ratio = 12]{\label{fig9d}\includegraphics[scale=0.35]
  {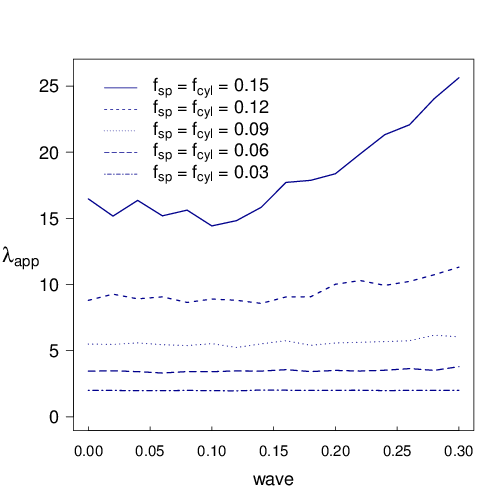}}
  \caption{The dependence of thermal conductivity of a composite material on the corrugating on the shape of inclusions. $n_{sp}= n_{cyl}=20$, contrast 2048}
  \label{fig9}
\end{figure}

The first case concerns the shape with corrugations and depicts the apparent thermal conductivity versus the wave parameter, i.e. the ratio between the amplitude of perturbations of the surface and the radius of the sphere and the cylinder. Fig. \ref{fig9} exhibits the influence of the wave parameter for several aspect ratios of cylinders (3, 6 and 9) and for contrast equal to 2048. 

For smaller contrast, the effect is less apparent, almost no modification is observed . We can observe that the apparent thermal conductivity  of these composite materials increase in comparison to a similar composite material with the same volume fraction of spheres and cylinders and the same aspect ratio. Moreover, we have the similar trends already observed with the mechanical properties. The effect is more pronounced for composite materials with volume fraction of inclusions higher than $20\%$ and for all aspect ratios tested.

The second series of tests concerns the composite materials with the second variation, an example is also shown on figure \ref{fig2b}.
For this series of tests, we have used the same generations of RVE but with different defects due to the random aspect of the generation of defects. Nevertheless, we observe the same trends as in the mechanical case (see \cite{Salnikov1}) that is $\lambda_{app}$ increases with the volume fraction of defects and with the aspect ratio. 

\begin{figure}[H]
  \centering
  \subfigure[Aspect ratio = 3]{\label{fig10a}\includegraphics[scale=0.35] {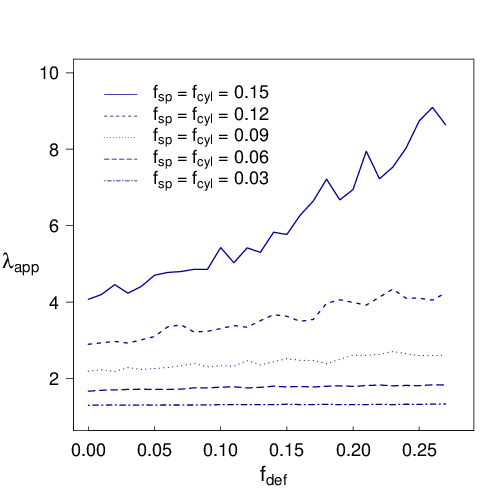}}
   \hspace{5pt}
  \subfigure[Aspect ratio = 6]{\label{fig10b}\includegraphics[scale=0.35] {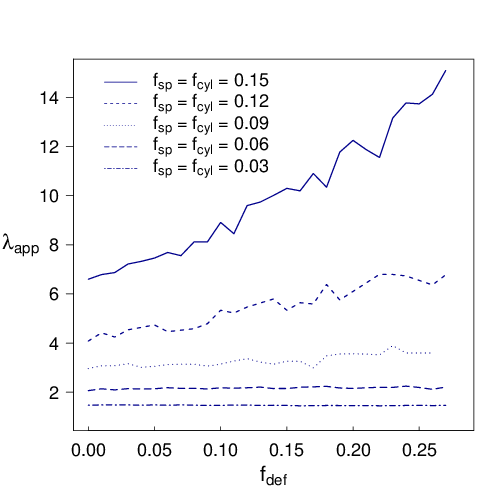}}
  \subfigure[Aspect ratio = 9]{\label{fig10c}\includegraphics[scale=0.35]
  {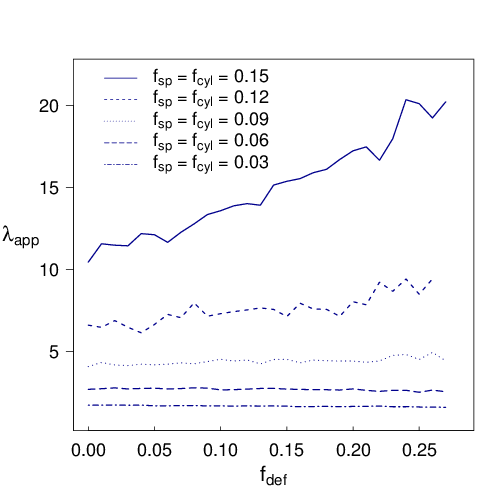}}
 \subfigure[Aspect ratio = 12]{\label{fig10d}\includegraphics[scale=0.35]
  {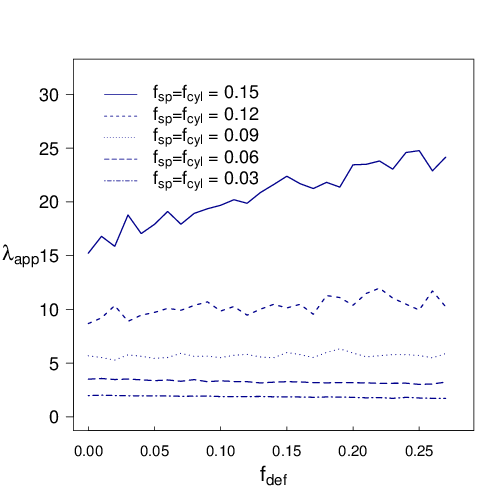}}
  \caption{The dependence of thermal conductivity of a composite material on the volume of defects in the material. $n_{sp}= n_{cyl}=20$, contrast 2048}
  \label{fig10}
\end{figure}

Precisely, we represent the results versus the value of volume fraction of defects and for different aspect ratio of cylinder with a contrast equal to 2048 (for the same reasons as it is mentioned previously) on Fig. \ref{fig10}. The higher the fraction of defects is, the higher the apparent thermal conductivity is. This observation is especially valid for composite materials with larger volume fraction and with cylinders with a large aspect ratio. We note that the effect observed with this series of test is more significant than the previous series.

\section{Comparison with mechanical properties - Conclusions and outlook}
\indent 

In this paper, we have recalled briefly the methods for the RVE generation composed by a mixture of spheres and cylinders. We have presented how adding some irregular shapes for inclusions in these RVEs. We have studied how a generated RVE which the mechanical behaviour previously studied would behave in a thermal point of view. Our work has been meticulous in order to use the same generated RVE. We have chosen a similar approach to make our computations. We use the 'FFT-based stochastic homogenization' by using the FFT-based iterative accelerated scheme to determine effective properties and the stochastic characteristic is obtained with the random generation of several RVEs with the same fixed parameters.

Finally, we observed qualitatively the same trends between the mechanical and the thermal computations. We can conclude that the morphological parameters influencing the thermal behaviour as the mechanical one are the volume fraction of inclusions and in particular the volume fraction of fibre type inclusions and the morphology of this kind of inclusions. We also conclude that adding some irregularities (waves and defects) improves the thermal behaviour and we have observed the same tendencies between the two series of irregularities tested and the two behaviours: mechanical and thermal. We also notice at large contrast that the higher is the aspect ratio, the greater is the observed effect.

This work provides a wide range of RVE with numerous parameters (volume fraction of spheres or cylinders, number of spheres or cylinders, morphology of cylinders with their aspect ratio, the irregularities as the shape with corrugations or cuts of inclusions). For each generated RVE the mechanical and thermal behaviour with a contrast varying from 0.0625 to 2048 had been computed.
We can imagine that the set of RVE generated can be an interesting database for research in inverse problems.

 The next step is to use a real composite material given by images generated by tomography and segmented in order to determine its effective thermal and mechanical properties. One of the objective of this study could be to compare the results obtained with generated RVEs with results obtained with the tomography images. Another point of our future work should take into account the large images size. Actually, images from tomography can be defined with high resolution (2672 $\times$ 2732 pixels per image for example) and the memory consumption becomes so high that it is not possible to make calculations directly. We have already made several tests with a sample but due to the high resolution of the images we must resort to a strategy using stochastic method. This work is in progress.

\section*{Acknowledgements}
\indent 

Most of the computations described in this paper have been carried out at the cluster of the Centre of Informatics and numeric applications of Normandy (CRIAAN - Centre Régional Informatique et d’Applications Numériques de Normandie).\\
This work has been supported by ACCEA project selected by the "Fonds Unique Interministériel (FUI) 15 (18/03/2013)" program.

 \begin{appendices}
 \section{Resume of our approach to determine the thermal properties}
 \label{approach}
\begin{enumerate}
\item Fix the macroscopic parameter of the material (volume fraction and type of inclusions).
\item Use the series of generated samples (RVEs) of a composite material with the fixed parameters (stochastic part).
\item Perform accurate computation of effective properties on these RVEs (deterministic homogenization part).
\item Average the computed macroscopic characteristics of the samples.
\end{enumerate}

\section{The thermal model}
\label{model}
Consider a representative volume element $V$ with periodic boundary conditions, and denote $\theta(\boldsymbol{x})$ the temperature, $\boldsymbol{\phi}(\boldsymbol{x})$ the heat flux and $L(\boldsymbol{x})$ the thermal conductivity at any point $\boldsymbol{x}  \in  V$. 

Fourier's law in the linear case states that: $\boldsymbol{\phi}(\boldsymbol{x})=-L(\boldsymbol{x})\: \nabla \theta(\boldsymbol{x})$. Moreover, we suppose that we are in stationary phase and without heat source so the heat flux satisfies $\mathop{\rm div}(\boldsymbol{\phi}(\boldsymbol{x})) = 0$. Notice that for a composite material, the thermal conductivity tensor depends on the point $\boldsymbol{x}$: the dependence is governed by microscopic geometry of the sample, namely which phase (matrix or inclusion) the point belongs to. 

The volume element $V$ is subjected to an average temperature gradient $<  \nabla \theta(\boldsymbol{x})>$ which induces local temperature gradient $\nabla \theta(\boldsymbol{x})$ and heat flux $\boldsymbol{\phi}(\boldsymbol{x})$  inside the RVE. The effective constitutive relations of the composite material are the relations between the (spatial) average flux $< \phi(\boldsymbol{x})>$ and average temperature gradient $<  \nabla \theta(\boldsymbol{x})>$ that is: 
$$< \phi(\boldsymbol{x})>=-L_{hom}< \nabla\theta(\boldsymbol{x})>.$$
To determine the components of $L_{hom}$ in 3-dimensional space, we need to compute $< \phi(\boldsymbol{x})>$ for three independent values of $< \nabla \theta(\boldsymbol{x})> \equiv \nabla \Theta$.

There are two approaches to recover $L_{hom}$: we can construct a mesh of the RVE V and use the finite elements method or discretize  (that is pixelize in 2D and voxelize in 3D) and use the FFT-based homogenization scheme (\cite{Moulinec1},\cite{Eyre and Milton}). We have chosen the latter method for its capability to take account of complex geometry with some computation reasonable memory although with the first method, we must make fine mesh for some cases of complex geometry which require a lot of memory. Moreover, we suppose the materials isotropic so that we have an exactly expression to the Green operator using in the Fourier method.  

More precisely, it is a common practice to introduce a constant reference tensor $\Lambda^{0}$ to solve the problem of recovering the local fields and gradients.
Introducing this homogeneous linear tensor and $ \delta L(\boldsymbol{x}) = (L(\boldsymbol{x})-\Lambda^{0})$, the problems reads:
\begin{equation}
\left \{
\begin{array}{l}
\boldsymbol{\phi}(\boldsymbol{x})=-\Lambda^{0}: \nabla \overset{\sim}{\theta}(\boldsymbol{x}) + \boldsymbol{\tau}(\boldsymbol{x}), ~ \forall \boldsymbol{x} \in V \\
\mathop{\rm div}(\boldsymbol{\phi}(\boldsymbol{x})) = 0, ~ \forall \boldsymbol{x} \in V \\
\boldsymbol{\tau}(\boldsymbol{x})= -\delta L(\boldsymbol{x}):(\nabla \overset{\sim}{\theta}(\boldsymbol{x}) +\nabla \Theta) - \Lambda^{0}: \nabla \Theta, ~ \forall \boldsymbol{x} \in V \\
\overset{\sim}{\theta}(\boldsymbol{x}) ~ \text{periodic,} ~ \boldsymbol{\phi}(\boldsymbol{x}).\boldsymbol{n} \text{ antiperiodic}
\end{array}
\right.
\end{equation}
where $\boldsymbol{\tau}$ denotes the polarization tensor.\\
The solution of (1) can be expressed in real and Fourier spaces, where $\widehat{\bullet}$ denotes the Fourier image of $\bullet$ and $\xi_{j}$'s are the coordinates in the Fourier space, respectively, with the help of the periodic Green operator $\Gamma^{0}$ associated with $\Lambda^{0}$ we also can write:\\
in real space:
\begin{equation}
\nabla \theta(\boldsymbol{x})= \Gamma^{0}*(\delta L(\boldsymbol{x}):\nabla\theta(\boldsymbol{x}))+\nabla\Theta ~~~ \forall \boldsymbol{x} \in V 
\end{equation}
or in Fourier space:
\begin{equation}
\widehat{\nabla \theta}(\boldsymbol{\xi})= \widehat{\Gamma}^{0}(\boldsymbol{\xi}):\widehat{(\delta L:\nabla\theta)} (\boldsymbol{\xi})~~~ \forall \boldsymbol{\xi} \neq 0 , ~ \widehat{\nabla \theta}(\boldsymbol{0})=\nabla \Theta
\end{equation}
The equations (2) and (3) are the Lippmann-Schwinger integral equations  in real and Fourier spaces respectively.
The Green operator $\widehat{\Gamma}^{0}$ is easily expressed and computed in the Fourier space by:
$$
\widehat{\Gamma}^{0}_{ij}(\boldsymbol{\xi})=\dfrac{\boldsymbol{\xi}_{i}\boldsymbol{\xi}_{j}}{\sum\limits_{m,n}\Lambda^{0}_{mn}\boldsymbol{\xi}_{m}\boldsymbol{\xi}_{n}}
$$

The Lippmann-Schwinger equation can be solved iteratively using the algorithm based on the accelerated scheme proposed by Eyre and Milton in \cite{Eyre and Milton} and written as algorithm 3 in \cite{Salnikov2} for the elastic case (see appendix \ref{algo}).

\section{Algorithm: FFT-based accelerated scheme}
\label{algo}
\textbf{Algorithm:} \textit{FFT-based homogenization scheme for thermal conductivity case}

\textit{initialize $\nabla \theta^{0}(\boldsymbol{x})\equiv <\nabla \theta(\boldsymbol{x})>\equiv \nabla \Theta$}\textit{ and fix the convergence criterion acc.}

\textit{while (not converged)}
      \newline \begin{tabular}{c|l}   & \parbox{0.98\linewidth}{
      \begin{enumerate}
       \item \textit{convergence test:}\\
       \textit{if (}$\epsilon_{comp} < acc)$ \textit{compute} $\boldsymbol{\phi}^{n}(\boldsymbol{x})=- L(\boldsymbol{x}):\nabla \theta^{n}(\boldsymbol{x})$, $\widehat{\boldsymbol{\phi}}^{n}= FFT(\boldsymbol{\phi}^{n})$,\\
       $\epsilon_{eq}=\sqrt{\langle\Vert\boldsymbol{\xi} \widehat{\boldsymbol{\phi}}^{n}(\boldsymbol{\xi})\Vert^{2}\rangle}/\Vert\widehat{\boldsymbol{\phi}}(\boldsymbol{0})\Vert$\\
       \textit{if} $(\epsilon_{eq} <acc)$ \textit{then the convergence is achieved and we stop}
       \item $\boldsymbol{\tau}^{n}(\boldsymbol{x})= -(L(\boldsymbol{x}) + \Lambda^{0}):\nabla \theta(\boldsymbol{x})$
       \item $\widehat{\boldsymbol{\tau}}^{n}= FFT(\boldsymbol{\tau}^{n})$
       \item $\widehat{\nabla \theta}^{n}_{comp}(\boldsymbol{\xi})=-\widehat{\Gamma}^{0}(\boldsymbol{\xi}):\widehat{\boldsymbol{\tau}}^{n}(\boldsymbol{\xi}), \boldsymbol{\xi} \neq 0$,  $\widehat{\nabla \theta}^{n}_{comp}(\boldsymbol{0})=\nabla \Theta$
       \item $\nabla \theta^{n}_{comp}= FFT^{-1}(\widehat{\nabla \theta}^{n}_{comp})$
       \item $\epsilon_{comp}=\sqrt{\langle\Vert \nabla \theta^{n}-\nabla \theta^{n}_{comp}\Vert^{2}\rangle} /\Vert \nabla \Theta\Vert$
       \item $\nabla \theta^{n+1}(\boldsymbol{x})= \nabla \theta^{n}(\boldsymbol{x}) - 2(L(\boldsymbol{x})-\Lambda^{0})^{-1}:\Lambda^{0}:(\nabla \theta^{n}_{comp}(\boldsymbol{x})-\nabla \theta^{n}(\boldsymbol{x}))$
       \end{enumerate}
        }\end{tabular}

\vspace{0.3cm}    
\noindent When this algorithm converges, we compute $< \boldsymbol{\phi}(\boldsymbol{x})>$ and we obtain $L_{hom}$ using the following equality: $< \boldsymbol{\phi}(\boldsymbol{x})> =-L_{hom}< \nabla \theta(\boldsymbol{x})>=-L_{hom} \nabla  \Theta $. As we have mentioned above, we repeat this computation for three independent values of $\nabla \Theta$.

\section{Notations and range of parameters with figure references}
\label{notations}
\renewcommand{\arraystretch}{1.3}
\begin{tabular}{>{\centering}p{1.5cm}|>{\centering}p{3.8cm}||>{\centering}p{2cm}|>{\centering}p{2cm}|>{\centering}p{2cm}}
notation & definition & Fig.\ref{fig3} & Fig.\ref{fig4}& Fig.\ref{fig5} \tabularnewline
\hline
$n_{sp}$& number of spheres & \small{$[0, \ 50]$} & \small{$\{15, \ 20, \ 25\}$} \tabularnewline
$n_{cyl}$&number of cylinders &  \small{$[0,\ 50]$} &  &\small{$20$} \tabularnewline
$n_{def}$& number of defects &  & & \tabularnewline
\hline
$f_{sp}$& fraction of spheres & \small{$[0.01, \ 0.15]$}& \small{$[0.01, \ 0.39]$} \tabularnewline
$f_{cyl}$& fraction of cylinders & \small{$[0.01, \  0.15]$} & &\small{$[0.01, \ 0.29]$} \tabularnewline
$f_{def}$& fraction of defects & & & \tabularnewline
\hline
$a$& aspect ratio for one cylinder (Fig. \ref{aspect}) & \small{$5$} &   & \small{$[1, \ 19]$} \tabularnewline
$wave$ &relative wave amplitude (Fig. \ref{wave}) & & &\tabularnewline
$contrast$ & contrast between matrix and inclusions& \small{$[2^{-4}, \  2^{11}]$}&\small{$[2^{-4}, \ 2^{11}]$}&\small{$[2^{-4}, \  2^{11}]$}\tabularnewline
\hline
\end{tabular}

\noindent
\renewcommand{\arraystretch}{1.3}
\begin{tabular}{>{\centering}p{1.5cm}||>{\centering}p{1.7cm}|>{\centering}p{2.4cm}|>{\centering}p{2cm}|>{\centering}p{2cm}|>{\centering}p{2cm}}
notation & Fig.\ref{fig6}& Fig.\ref{fig7}&Fig.\ref{fig8}&Fig.\ref{fig9}& Fig. \ref{fig10} \tabularnewline
\hline
$n_{sp}$ & \small{$20$} & \small{$[6, \ 27]$} &\small{$20$}&\small{$20$}&\small{$20$}\tabularnewline
$n_{cyl}$ & \small{$20$}  & \small{$[6,\ 27]$} &\small{$20$}&\small{$20$}&\small{$20$}\tabularnewline
$n_{def}$ & & & & &\small{$30$}   \tabularnewline
\hline
$f_{sp}$ & \small{$[0, \ 0.15]$}&\small{$\{0.05, 0.1, 0.15\}$}& \small{$[0.01, \  0.15]$}&\small{$[0.03, \  0.15]$}&\small{$[0.03 , \ 0.15]$}\tabularnewline
$f_{cyl}$ & \small{$[0, \ 0.15]$} &\small{$\{0.05,0.1,0.15\}$}&\small{$[0.01, \ 0.15]$}&\small{$[0.03, \ 0.15]$}&\small{$[0.03, \ 0.15]$}\tabularnewline
$f_{def}$ & & & & & \small{$[0, \  0.27]$}\tabularnewline
\hline
$a$&  \small{$[1, \ 19]$}  & \small{$5$} &\small{$5$} &\small{$\{3,\ 6,\ 9,\ 12\}$}&\small{$\{3,\ 6,\ 9,\ 12\}$}\tabularnewline
$wave$ & & & &\small{$[0, \ 0.3]$}& \tabularnewline
$contrast$ & \small{$[2^{-4}, \ 2^{11}]$}&\small{$[2^{-4}, \ 2^{11}]$}&\small{$[2^{-4}, \ 2^{11}]$}&\small{$[2^{-4}, \ 2^{11}]$}&\small{$[2^{-4}, \ 2^{11}]$}\tabularnewline
\hline
\end{tabular}

\end{appendices}



\begin{thebibliography}{9}
\bibitem{Salnikov1}
 V. Salnikov, S. Lemaitre, D. Choi, Ph. Karamian-Surville
Measure of combined effects of morphological parameters of inclusions within composite materials via stochastic homogenization to determine effective mechanical properties, 
Composite Structures Volume 129 pp 122-131, October 2015


\bibitem{segurado}
J. Segurado, J. Llorca, A numerical approximation to the elastic properties of sphere-reinforced composites, Journal of the Mechanics and Physics of Solids, 50 (2002) 2107-2121

\bibitem{ghossein}
E. Ghossein, M. Lévesque, A fully automated numerical tool for a comprehensive validation of homogenization models and its application to spherical particles reinforced composites, International Journal of Solids and Structures 49, (2012) 1387-1398

\bibitem{zhao}
J. Zhao, S. Li, R. Zou, A. Yu, Dense random packings of spherocylinders, Soft Matter, 8 (2012) 10031009

\bibitem{man}
W. Man, A. Donev, F. Stillinger, M. Sullivan, W. Russel, D. Heeger, S. Inati, S. Torquato, P. Chaikin, Experiments on random packings of ellipsoids, Physical Review Letters, 94, (2005) 198001



\bibitem{Salnikov2}
 V. Salnikov, D. Choi, Ph. Karamian-Surville
On efficient and reliable stochastic generation of RVEs for analysis of composites within the framework of homogenization, 
Computational Mechanics, Volume 55, Issue 1, 2015. 

\bibitem{Eyre and Milton}
 D.J. Eyre, G.W. Milton 
A fast numerical scheme for computing the response of composites using grid refinement, Journal of Physique III 1999; 6; 41-47

\bibitem{Moulinec1}
 H. Moulinec and P. Suquet, A fast numerical method for computing the linear and non linear properties of composites, C.R Acad. Sc. Paris II 318 (1994) 1417-1423

\bibitem{CSMA}
 S. Lemaitre, V. Salnikov, D. Choi, P. Karamian
Approche par la dynamique moléculaire pour la conception de VER 3D et variations autour de la pixellisation, accepted for publication in the proceedings of the CSMA 2015 




\bibitem{Moulinec and Silva}
 H. Moulinec, F. Silva
Comparison of the three accelerated FFT-based schemes for computing the mechanical response of composite materials, Int. J. Numer. Meth. Engng. 2014; 97:960-985

\bibitem{Moulinec and Suquet 2001}
 J.C. Michel, H. Moulinec,  P. Suquet
A computational scheme for linar and non-linear composites with arbitrary phase contrast , Int. J. Numer. Meth. Engng. 2001; 52:139-160(DOI: 10.1002/nme.275)

\end{thebibliography}
\end{document}